# Heterostructures of hetero-stack of 2D TMDs ($MoS_2$, $WS_2$ and $ReS_2$) and BN

Badri Vishal,[1] U. Bhat,[1] H. Sharona,[1] and Ranjan Datta[1,*]

[1]*International Center for Materials Science, Chemistry and Physics of Materials Unit, Jawaharlal Nehru Center for Advanced Scientific Research, Bangalore 560064.*

In this manuscript, we describe optical emission of heterostructure of hetero-stack between 2D TMDs ($MoS_2$, $WS_2$, and $ReS_2$) and BN. Similar to our previous results on the stack of similar type of TMDs, intense PL emission peak is observed around 2.13 eV but is split around 2.13eV into two or more peaks depending on the different stack of TMDs with BN. The transitions from the valence band of BN to conduction bands of different TMD stacks due to quantum coupling and specific orientation explain the strong peak in the PL spectra.

*Corresponding author e-mail: ranjan@jncasr.ac.in*

# 1. Introduction

2D-transition metal dichalcogenides (TMDs) exhibits various unique electronic properties in monolayer thickness [1-4]. A combination of these materials among themselves and with insulators like boron nitride (BN) either in-plane or vertical, known as van der Waals (vdW) compounds, may offer an immense possibility of tuning the existing properties of individual TMDs and exploring new physical phenomena [5-11]. The vdW heterostructures show a significant increase in mobility in graphene on the *h*-BN substrate [12]. *h*-BN substrate increases the mobility of $MoS_2$, $WSe_2$ significantly [13]. Vertical heterostructure of $MoS_2$-$WSe_2$ has been demonstrated to be a thinnest *p-n* junction [14]. $MoS_2$ and *h*-BN form type-I heterostructure, which is suitable for optoelectronic devices like laser and light-emitting diodes. Most of the TMDs form type-I heterostructure with *h*-BN [15-19].

Arranging such layers of 2D-materials in heterostructure form is challenging. Initially, it was carried out by isolating individual layers on a thin transparent polymer film and sticking them face to face [20]. The exfoliation method is prone to contamination and not suitable for large area device applications. Therefore, numerous groups explored alternative methods for the growth of wafer-scale vdW heterostructures, e.g., chemical vapor deposition (CVD), pulsed laser deposition (PLD), atomic layer deposition (ALD), etc. [6,19-26].

In the context, PLD has already been used under the slow kinetic condition to grow large area (10×10 mm$^2$) epitaxial films of $MoS_2$, $WS_2$, $ReS_2$, and their heterostructures with BN on *c*-plane sapphire substrate [19,24-26]. The heterostructure systems thus have grown between TMDs and BN show strong emission around 2.3 eV with minor subsidiary peaks. The observation of strong emission at 2.3 eV was explained from the VB of BN to CB of various TMDs after considering relative orientation difference of 30° between the hexagonal lattice of TMDs and BN. This changes the $\Gamma(BN) \rightarrow K(\text{TMDs})$ transition as direct optical transition channel. In the present contribution, the extension of the above results based on heterostructures of different TMDs with BN in the same stack is demonstrated. However, in the present case, the most intense peak around 2.13 eV is split into two and three different peaks depending on the number of different TMDs in the hetero-stack. This suggests the coupling of the BN valence band to three different conduction bands of TMDs under specific stacking orientation.

## 2. Experimental methods

Heterostructures of $MoS_2$ (or $WS_2$ or $ReS_2$)/BN/$MoS_2$(or $WS_2$ or $ReS_2$)/BN were grown by the PLD technique under slow kinetic conditions as described earlier[19,24-26]. $MoS_2$, $WS_2$, and *h*-BN target pellets were prepared from powders obtained from Sigma Aldrich (99.9% purity) by first cold pressing and then sintering at 500 °C for 5 hours in a vacuum chamber (~$10^{-5}$ Torr). Sintering in the vacuum chamber prevents oxidation of compounds as well as re-deposition of vapor species back on the pellet surface, unlike sintering performed in a sealed quartz tube. The final growth scheme for three different types of Hetero-stack of vdW heterostructure on the sapphire substrate by PLD is shown in Supplementary Figure S1. Hetero-stacks of $MoS_2$/BN/$WS_2$/BN, $MoS_2$/BN/$ReS_2$/BN, and $MoS_2$/BN/$WS_2$/BN/ $ReS_2$/BN are grown on the *c*-plane sapphire substrate. External interactions are minimized by encapsulating the heterostructure with *h*-BN.

PL measurement was performed with 510 nm Green Laser diode excitation in LabRAM HR (UV) system. Raman spectra were recorded using a custom-built Raman spectrometer using a 514.5nm laser excitation and a grating of 1800 lines/mm at room temperature. The laser power at the sample was approximately 1mW [27].

## 3. Results and discussion

Figure 1. is the schematic of three different heterostructure of hetero-stack between TMDs and BN. In the hetero-stack system, the first layer is always MoS$_2$. In Figure 1 (a) and (b), the top TMD layers are WS$_2$ and ReS$_2$ respectively. In Figure 1(c) the hetero-stack consisting of three different TMDs, the final TMD layer is ReS$_2$. It is found that the crystal structure of BN is different depending on the TMD template used. BN tends to grow as *h*-BN on top of MoS$_2$, whereas it forms a mixture of *c*-BN and *w*-BN on top of WS$_2$ and ReS$_2$ [26]. Raman spectra of (a) MoS$_2$/BN (b) WS$_2$/BN (c) ReS$_2$/BN heterostructure grown on top of the sapphire substrate are given in Figure 2. In Figure 2(a) characteristic A$_{1g}$ and E$^1_{2g}$ modes located at 380.1 and 406.2 cm$^{-1}$ corresponding to MoS$_2$ are indicated. E$^1_{2g}$ mode is due to the vibration of two S atoms in the opposite direction, while A$_{1g}$ mode is due to the out-of-plane vibration of S atoms in opposite directions to the Mo atom [28]. The Raman peak at 751 and 1372 cm$^{-1}$ belong to A$_{2u}$ (TO) and E$_{2g}$ of *h*-BN [29]. In Figure 2(b), E$^1_{2g}$ and A$_{1g}$ are characteristic modes located at 354 and 417 cm$^{-1}$ corresponds to WS$_2$ are indicated. The peaks at 750, 835, 1373 and 1600 cm$^{-1}$ corresponding to A$_{2u}$ (TO), A$_{2u}$ (LO), E$_{2g}$ and E$_{1u}$ (LO) of h-BN grown on top of WS$_2$ are also indicated. The characteristic modes corresponding to ReS$_2$ is indicated in Figure 2(c). For ReS$_2$ E$_g$ modes at 165cm$^{-1}$ are due to in-plane vibrations of Re atoms, A$_g$ modes located at 420 and 440 cm$^{-1}$ are from out-of-plane vibrations of S atoms and the C$_p$ modes at 275 cm$^{-1}$ are due to in and out-of-plane vibration of Re and S atoms, while 242 cm$^{-1}$ is in-plane vibration of Re atoms, and 407.3 cm$^{-1}$ is the in-plane and out-of-plane vibration of S atoms [30,31]. Both *w*-BN and *h*-BN have been identified by Raman spectroscopy; 970 and 985 cm$^{-1}$ are E$_2$ and B$_1$ of *w*-BN, while 1353 and 1612 cm$^{-1}$ are E$_{2g}$ and E$_{1u}$ (LO) of *h*-BN [29].

The photoluminescence (PL) spectra at room temperature for three different heterostructure systems, $MoS_2$/BN/$WS_2$/BN, $MoS_2$/BN/$ReS_2$/BN, and $MoS_2$/BN/$WS_2$/BN/$ReS_2$/BN are given in Figure 3. For stack consisting of $MoS_2$/BN/$WS_2$/BN, two intense peaks at 2.16 and 2.13 eV can be observed other than subsidiary peaks around 2 and 1.97 eVs. Similar splitting in the intense peak is observed for stack consisting of $MoS_2$/BN/$ReS_2$/BN. The splitting becomes systematically three for the stack of $MoS_2$/BN/$WS_2$/BN/$ReS_2$/BN, consisting of three different TMDs. Figure 4. is the schematic of the band alignment to explain the origin of all the observed emission peaks. The most intense peak is around 2.13 (±0.03) eV is common in all the hetero-stacks. This is similar to our previous work [19] where the intense peak is observed at 2.3 eV. In the present case, the mixture of different TMDs affecting the intense peak to split into multiple peaks depending on the number of TMDs present in the hetero-stack. Band alignment with respect to vacuum does not differ much from each other for these TMDs [32] which explains similar peaks for different hetero-stack systems at 2.13 eV. Comparatively smaller peaks around 1.9 and 2eV are due to A and B excitons in the bulk $MoS_2$ [33,34]. The difference in the energy can be because of the defects or impurities present in the systems.

As we could not verify the orientation of BN on TMDs with experimental data, we explain the most intense peak at 2.13 eV based on available theoretical bandstructure calculations. Heterostructures of TMDs with *h*-BN tend to behave in a different way than when they are separate. Band structures of such systems are given in supplementary figure S2 for reference. Figure S2(a) is the band structure of $MoS_2$/*h*-BN and (b) $WS_2$/*h*-BN [35,36]. Considering the example of $MoS_2$/*h*-BN, the indirect transition from VB of *h*-BN to CB of $MoS_2$ (K $\longrightarrow$ *M*) results in a peak of around 2.2eV in the PL spectra which may not be intense. This is because the calculations presented in reference 35 and 36 were carried out considering a particular orientation where Brillouin zones of TMD, as well as h-BN were aligned in the same direction. Since both are vdW layer materials, quantum mechanical interactions between the two layers will be weak. This affects the orientation of BN layers on top of TMDs. During the formation of the heterostructures, BN layers are rotated to 30º with respect to TMDs, i.e. M-point of BN coincides with the K-point of $MoS_2$ [19]. Now the transition from M → K, which is direct, results in an intense peak at 2.2 eV. There is a difference in the theory and experimental results since the calculations in density functional theory (DFT) only considers the ground state energy and also it is at absolute temperature. Also, it is only for one type of heterostructure, but what we have here is the stack of TMDs. In the case of experimental data, defects present in the systems, impurities may contribute to the same.

## 4. Conclusion

In conclusion, large-area growth of heterostructures of three different TMD and BN by PLD is demonstrated. Raman spectra show BN can form with different crystal structures depending on the type of TMD as a template layer. PL spectra of hetero-stack give PL response near 2.13 and 2.16 eV from each stack of TMD, which extends the possibilities to optoelectronic devices of hetero-stacks. The origin of the PL emission can be explained by considering the relative orientation of the band structure of hetero-stack systems. The PL response suggests all vdW heterostructure are optically active material and maybe a good candidate for solar photovoltaic application.

**Acknowledgments**

The authors at JNCASR sincerely acknowledge ICMS for the funding.

**FIGURES**

**Figure 1**

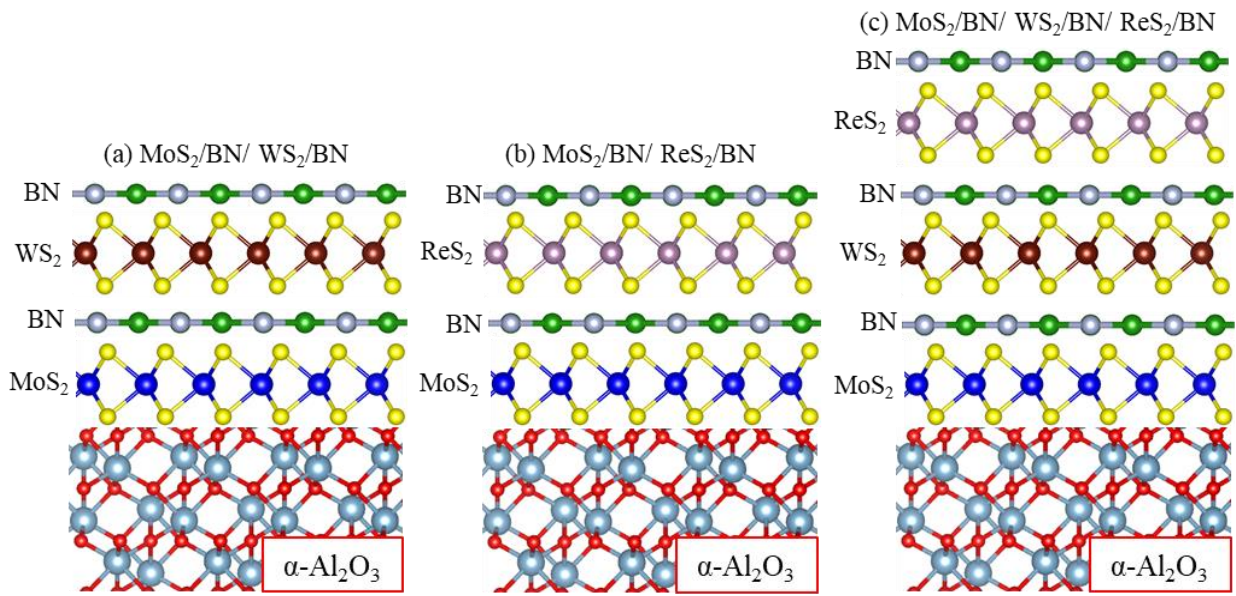

Figure 1. Schematic of van der Waals heterostructure of hetero-stack (a) $MoS_2/BN/WS_2/BN$

(b) $MoS_2/BN/ReS_2/BN$ (c) $MoS_2/BN/WS_2/BN/ReS_2/BN$

**Figure 2**

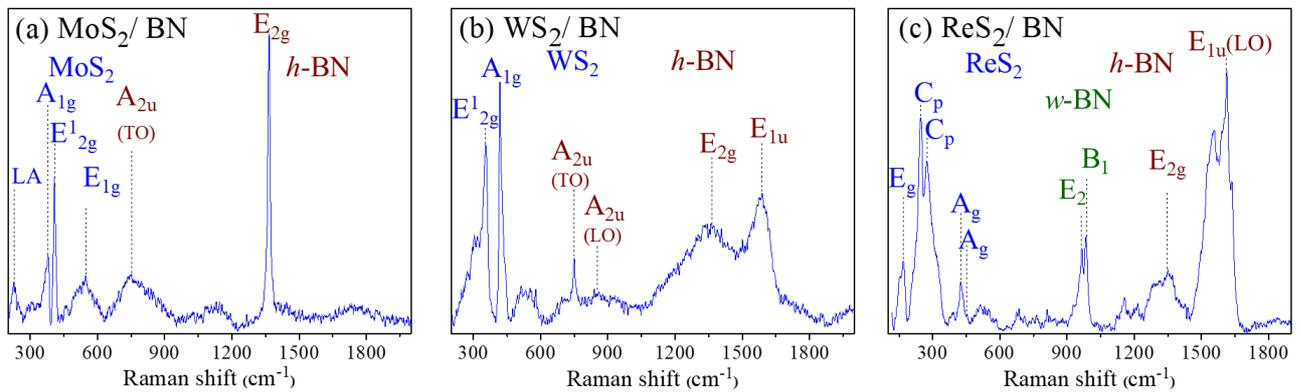

Figure 2. Raman spectra of (a) $MoS_2/BN$ (b) $WS_2/BN$ and (c) $ReS_2/BN$ heterostructures.

**Figure 3**

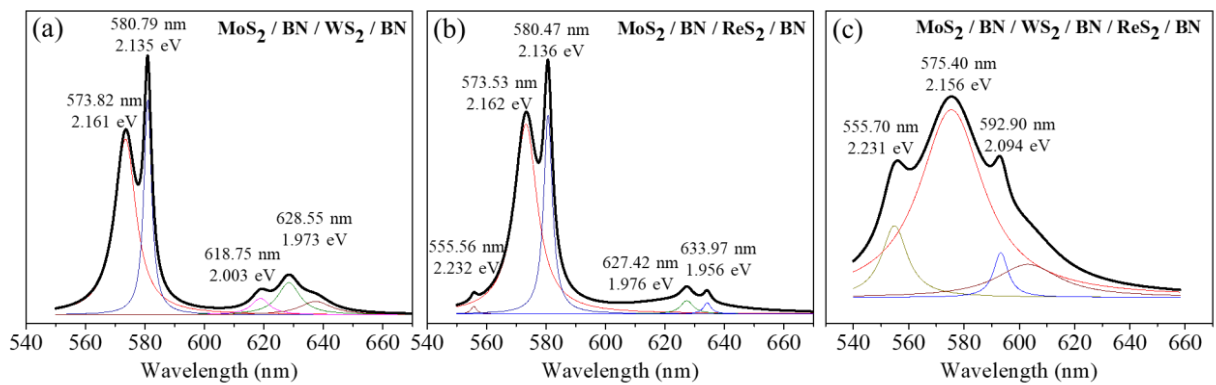

Figure 3. Photoluminescence spectra of three different heterostructure systems: (a) $MoS_2/BN/WS_2/BN$, (b) $MoS_2/BN/ReS_2/BN$, and (c) $MoS_2/BN/WS_2/BN/ReS_2/BN$.

**Figure 4**

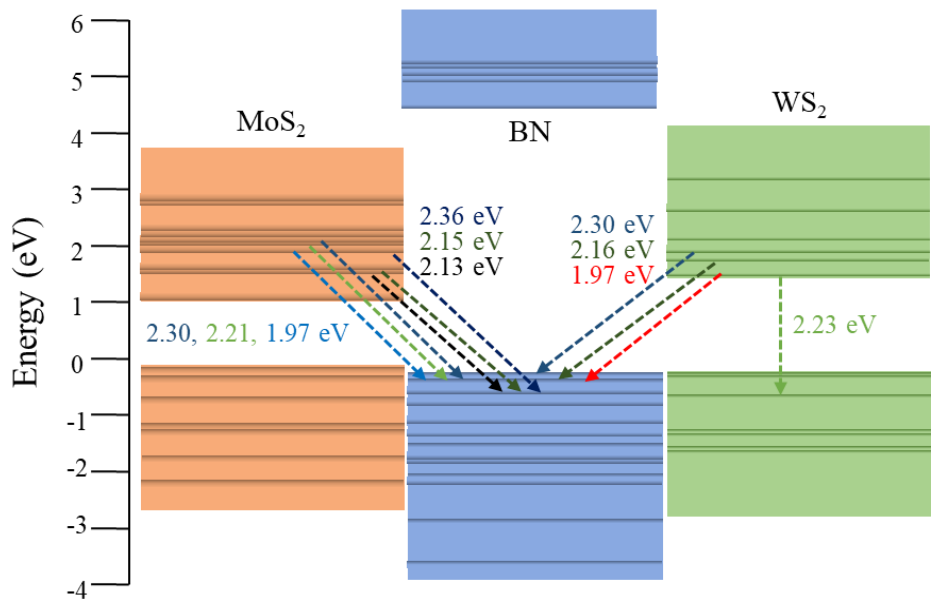

Figure 4. Schematic band to band transitions.

# Supplementary material

# Heterostructures of hetero-stack of 2D TMDs (MoS₂, WS₂, and ReS₂) and BN

**Figure S1:** Growth scheme for three different types of Hetero-stack of vdW heterostructures

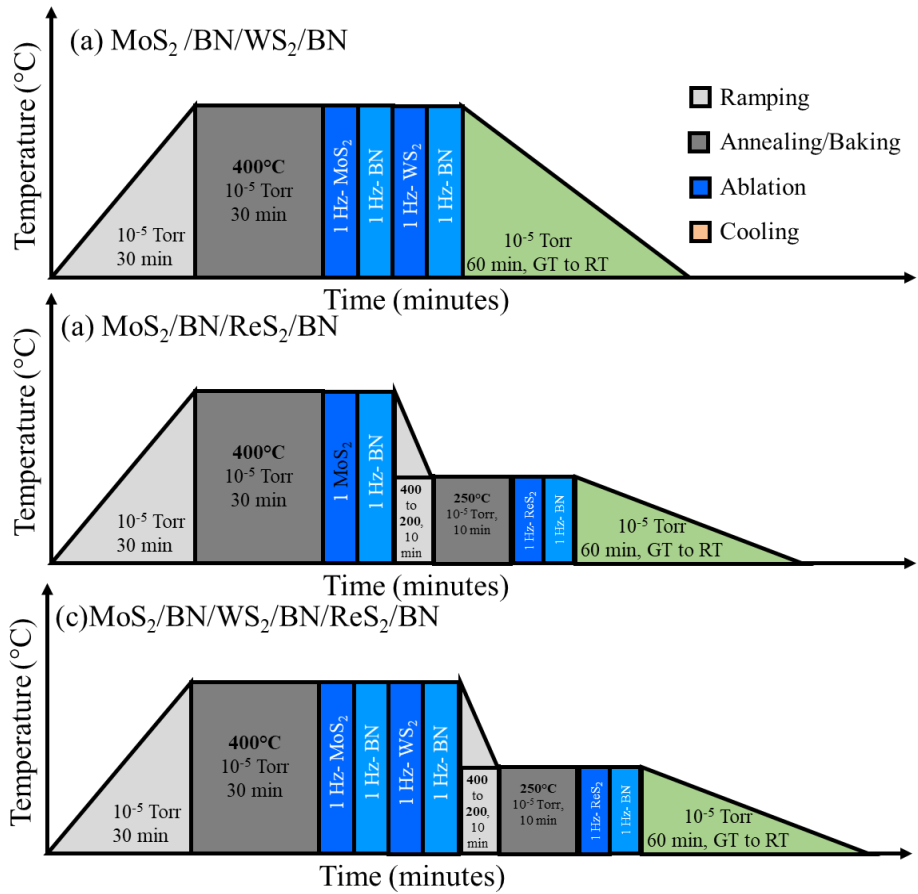

**Figure S1.** Schematic of growth scheme for three different types of Hetero-stack of vdW heterostructures on the sapphire substrate by PLD.

**Figure S2: Band structure of MoS₂/*h*-BN and WS₂/*h*-BN**

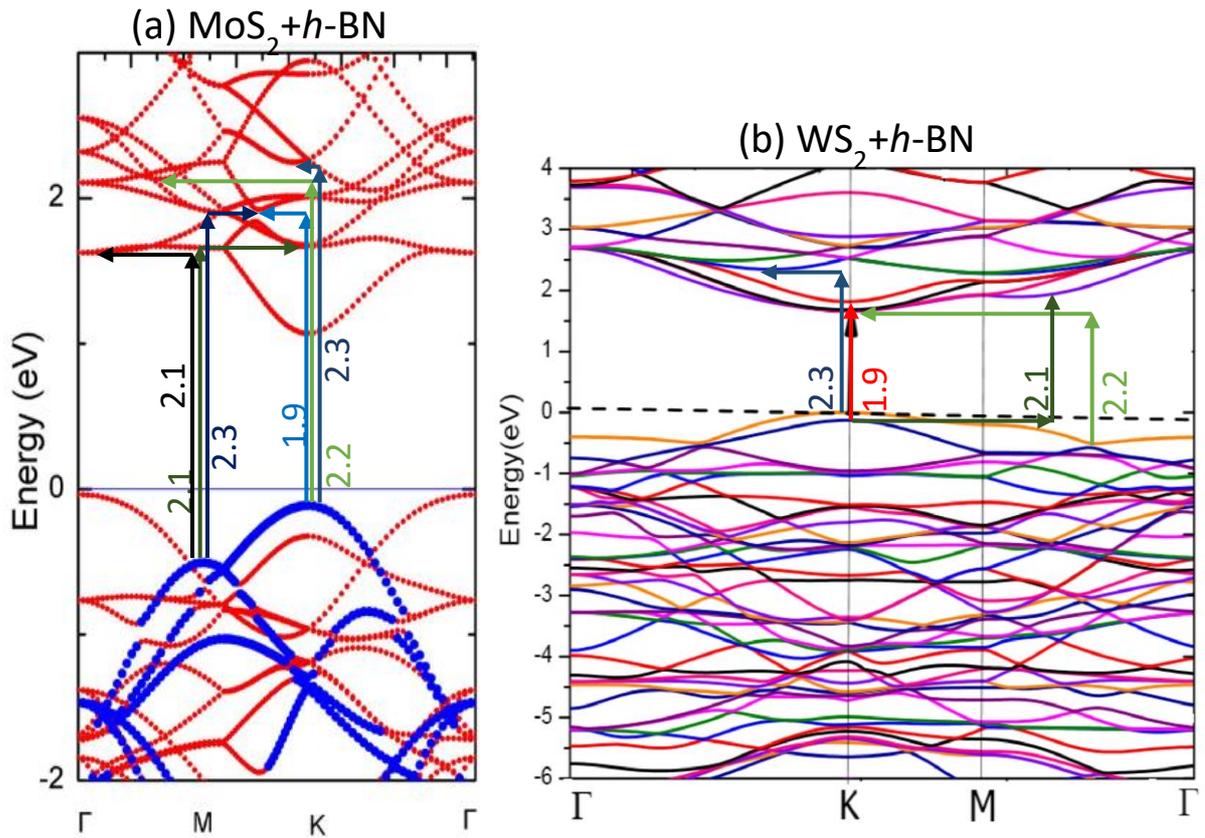

**Figure S2.** Band structure of (a) MoS2 and *h*-BN from reference 35 (b) WS2 and *h*-BN from reference 36.

As We Can see from band structure the emission of 1.97, 2.13, 2.15, 2.21, 2.30, 2.36 eV is possible for heterostructure between MoS2 and *h*-BN. Similarly, in the case of WS2 and *h*-BN band structure 1.97, 2.16, 2.23 2.30 eV is possible.